\newif\ifsingle
\singlefalse
\ifsingle
\documentclass[12pt, journal, onecolumn, table, draftclsnofoot]{IEEEtran}
\else
\documentclass[conference,10pt,table]{IEEEtran}
\fi

\ifCLASSOPTIONcompsoc
    \usepackage[caption=false, font=normalsize, labelfont=sf, textfont=sf]{subfig}
\else
\usepackage[caption=false, font=footnotesize]{subfig}
\fi
\usepackage[utf8]{inputenc}
\usepackage{lipsum}
\usepackage{array} 
\usepackage{epsfig}
\usepackage{epstopdf}
\usepackage{times}
\usepackage{framed}
\usepackage{algorithm}
\usepackage{enumitem}
\usepackage{float}
\usepackage{etoolbox}
\usepackage{algpseudocode}
\usepackage{amssymb}
\usepackage{amsmath}
\usepackage{multirow}
\usepackage{url}
\usepackage{lipsum}
\usepackage{soul}
\usepackage{caption}
\usepackage{mathtools}
\usepackage{bbm}
\usepackage{booktabs}
\usepackage{xcolor}
\usepackage{makecell}
\usepackage{pifont}
\usepackage{tabularx}
\usepackage[export]{adjustbox} 
\usepackage{latexsym}
\usepackage{cite}
\usepackage[acronym]{glossaries}
\usepackage{multicol}
\usepackage{eurosym}
\usepackage{xspace}
\usepackage{pdfpages}
\usepackage{verbatim}
\usepackage{stfloats}
\usepackage{mathtools}

\usepackage{blindtext}

\newacronym{ula}{ULA}{uniform linear array}
\newacronym{ga}{GA}{genetic algorithm}
\newacronym{csit}{CSIT}{channel state information at the transmitter}
\newacronym{cdf}{CDF}{cumulative distribution function}
\newacronym[plural=BSs, firstplural=base stations (BSs)]{bs}{BS}{base station}
\newacronym{pdf}{pdf}{probability distribution function}
\newacronym{aod}{AoD}{angle of departure}
\newacronym{aoa}{AoA}{angle of arrival}
\newacronym{ue}{UE}{user equipment}
\newacronym{los}{LoS}{line-of-sight}
\newacronym{pla}{PLA}{planar linear array}
\newacronym[plural=RISs, firstplural=reconfigurable intelligent surfaces (RISs)]{ris}{RIS}{reconfigurable intelligent surface}
\newacronym{sdp}{SDP}{semidefinite programming}
\newacronym{sdr}{SDR}{semidefinite relaxation}
\newacronym{svd}{SVD}{singular value decomposition}
\newacronym{slnr}{SLNR}{signal-to-leakage-and-noise ratio}
\newacronym{sre}{SRE}{smart radio environment}
\newacronym{snr}{SNR}{signal-to-noise ratio}
\newacronym{toa}{ToA}{time-of-arrival}
\newacronym{doa}{DoA}{direction-of-arrival}
\newacronym{mmse}{MMSE}{minimum mean squared error}
\newacronym{mimo}{MIMO}{multiple-input multiple-output}
\newacronym{miso}{MISO}{multiple-input single-output}
\newacronym{peb}{PEB}{position error bound}
\newacronym{oeb}{OEB}{orientation error bound}
\newacronym{rss}{RSS}{received signal strength}
\newacronym{ml}{ML}{machine learning}
\newacronym{rmse}{RMSE}{root-mean-square error}
\newacronym{mmwave}{mm-Wave}{millimeter-wave}
\newacronym{csi}{CSI}{channel state information}
\newacronym{3gpp}{3GPP}{3rd Generation Partnership Project}
\newacronym{sinr}{SINR}{signal-to-interference-plus-noise ratio}
\newacronym{smse}{SMSE}{sum mean squared error}
\newacronym{sbr}{SBR}{shooting and bouncing rays}
\newacronym{ura}{URA}{uniform rectangular array}
\newacronym{ofdma}{OFDMA}{orthogonal frequency-division multiple access}
\newacronym[plural=CSs, firstplural=candidate sites (CSs)]{cs}{CS}{candidate site}
\newacronym{mrt}{MRT}{maximum ratio transmission}
\newacronym{bca}{BCA}{Block Coordinate Ascent}
\newacronym{fp}{FP}{Fractional Programming}
\newacronym{tdma}{TDMA}{time-division multiple access}
\newacronym{jfi}{JFI}{Jain's fairness index}
\newacronym[plural=UAVs, firstplural=unmanned aerial vehicles (UAVs)]{uav}{UAV}{unmanned aerial vehicle}
\newacronym{ios}{IoS}{Internet of Surfaces}
\newacronym{soa}{SoA}{state-of-the-art}
\newacronym{rf}{RF}{radio frequency}
\newacronym{wmmse}{wMMSE}{weighted MMSE}
\newacronym{bcd}{BCD}{block coordinate descent}
\newacronym[plural=ESOs, firstplural=environmental scattering objects (ESOs)]{eso}{ESO}{environmental scattering object}

\newacronym[plural=RATs, firstplural=radio access technologies (RATs)]{rat}{RAT}{radio access technologies}

\newcommand{\rmscat}{\scriptscriptstyle\mathrm{SC}}

\newcommand{\rme}{\scriptscriptstyle\mathrm{E2E}}
\newcommand{\rmE}{\scriptscriptstyle\mathrm{E}}
\newcommand{\rmRIS}{\scriptscriptstyle\mathrm{RIS}}

\newcommand{\rmT}{\scriptscriptstyle\mathrm{T}}
\newcommand{\rmR}{\scriptscriptstyle\mathrm{R}}
\newcommand{\rmS}{\scriptscriptstyle\mathrm{S}}
\newcommand{\rmd}{\scriptscriptstyle\mathrm{d}}
\newcommand{\rmL}{\scriptscriptstyle\mathrm{L}}

\newcommand{\rmO}{\scriptscriptstyle\mathrm{O}}

\newcommand{\rmG}{\scriptscriptstyle\mathrm{G}}

\newcommand{\rmUS}{\scriptscriptstyle\mathrm{US}}
\newcommand{\name}{SARIS}

\newcommand{\change}[1]{{{#1}}}
\newcommand{\rmris}{\textnormal{\tiny{RIS}}}
\newcommand{\rmue}{\textnormal{\tiny{UE}}}
\newcommand{\rmbs}{\textnormal{\tiny{BS}}}

\IEEEoverridecommandlockouts

\DeclarePairedDelimiter\abs{\lvert}{\rvert}%
\DeclarePairedDelimiter\norm{\lVert}{\rVert}%
\newcommand{\overbar}[1]{\mkern 1.5mu\overline{\mkern-1.5mu#1\mkern-1.5mu}\mkern 1.5mu}

\makeatletter
\let\oldabs\abs
\def\abs{\@ifstar{\oldabs}{\oldabs*}}

\let\oldnorm\norm
\def\norm{\@ifstar{\oldnorm}{\oldnorm*}}
\makeatother

\renewcommand{\b}{\mathbf{b}}

\newcommand{\h}{\mathbf{h}}

\newcommand{\n}{\mathbf{n}}

\newcommand{\p}{\mathbf{p}}

\newcommand{\s}{\mathbf{s}}

\newcommand{\w}{\mathbf{w}}

\newcommand{\z}{\mathbf{z}}

\newcommand{\0}{\mathbf{0}}

\newcommand{\G}{\mathbf{G}}
\renewcommand{\H}{\mathbf{H}}
\newcommand{\I}{\mathbf{I}}

\newcommand{\W}{\mathbf{W}}

\newcommand{\Z}{\mathbf{Z}}

\newcommand{\deltab}{\boldsymbol{\delta}}

\newcommand{\mub}{\boldsymbol{\mu}}

\newcommand{\Deltab}{\mathbf{\Delta}}

\newcommand{\Compl}{\mbox{$\mathbb{C}$}}

\newcommand{\Exp}{\mathbb{E}}

\newcommand{\herm}{\mathrm{H}}
\renewcommand{\Im}{\mathrm{Im}}

\renewcommand{\Re}{\mathrm{Re}}
\newcommand{\tr}{\mathrm{tr}}
\newcommand{\tran}{\mathrm{T}}

\begin{document}
\bstctlcite{IEEEexample:BSTcontrol}
\ifsingle
\else
\title{\change{\huge SARIS: Scattering Aware Reconfigurable Intelligent Surface model and Optimization for Complex Propagation Channels}}
\fi

\ifsingle
\else
    \author{Placido Mursia,~\IEEEmembership{Member,~IEEE,} Sendy Phang,~\IEEEmembership{Member,~IEEE,} Vincenzo Sciancalepore,\\\IEEEmembership{Senior Member,~IEEE,} Gabriele Gradoni,~\IEEEmembership{Member,~IEEE,} and Marco Di Renzo,~\IEEEmembership{Fellow,~IEEE} \vspace{-0.5cm}
    \thanks{P. Mursia (corresponding author) and V. Sciancalepore are with NEC Laboratories Europe, 69115 Heidelberg, Germany (email: name.surname@neclab.eu).}
  	\thanks{S. Phang and G. Gradoni are with University of Nottingham, University Park, Nottingham, NG7 2RD, UK (email: name.surname@nottingham.ac.uk).}
  	\thanks{M. Di Renzo is with Université Paris-Saclay, CNRS, CentraleSupélec, Laboratoire des Signaux et Systèmes, 3 Rue Joliot-Curie, 91192 Gif-sur-Yvette, France (email: marco.di-renzo@universite-paris-saclay.fr).}
  	\thanks{This work was supported in part by EU H2020 projects RISE-6G and ARIADNE (grant agreements 101017011 and 871464, respectively). The research work of M. Di Renzo and G. Gradoni was also supported by the Fulbright Foundation and the Royal Society Industry Fellowship (grant number INF\textbackslash R2\textbackslash192066), respectively.}
 }
\fi

\maketitle

\begin{abstract}

The \gls{ris} is an emerging technology that changes how wireless networks are perceived, therefore its potential benefits and applications are currently under intense research and investigation. In this letter, we focus on electromagnetically consistent models for \glspl{ris} \change{inheriting from} a recently proposed model based on mutually coupled loaded wire dipoles. While existing related research focuses on free-space wireless channels \change{thereby ignoring interactions between \gls{ris} and scattering objects present} in the propagation environment, we \change{introduce} an \gls{ris}-aided channel model that is applicable to more realistic scenarios, \change{where the scattering objects are} modeled as loaded wire dipoles. By adjusting the parameters of the wire dipoles, the properties of general natural and engineered material objects can be modeled. \change{Based on this model}, we introduce a provably convergent and efficient iterative algorithm that jointly optimizes the \gls{ris} and transmitter configurations to maximize the system sum-rate. Extensive numerical results show the net performance improvement provided by the proposed method compared with existing optimization algorithms.


\end{abstract}
\glsresetall
\section{Introduction}

\change{The \gls{ris} technology is gaining growing attention in academia and industry owing to its ability to turn the stochastic nature of the wireless propagation channel---always conceived as a black-box---into an optimizable variable~\cite{Strinati2021}. \glspl{ris} are dynamic engineered metasurfaces, which are made of inexpensive scattering elements (unit cells) and low-complex/limited-power electronic circuits, which are spaced at sub-wavelength distances: the essentials are their dynamic reconfigurability in terms of scattering parameters performed in a nearly passive manner.
In this context, adequate communication and channel models for RISs are of paramount importance to obtain a deep understanding of their achievable performance and to fully exploit their potential in future wireless networks~\cite{DBLP:journals/pieee/RenzoDT22}. 

{\bf Related work.} In sub-wavelength implementations, there is an impelling need of considering the mutual coupling interactions among the unit cells: the discrete dipole approximation (DDA)---which is an analytical method for modeling the electromagnetic scattering from arbitrarily shaped objects~\cite{YURKIN2007558, 504309}---comes to help. In particular, the authors of~\cite{Gra21} have recently introduced a mutual coupling-aware and electromagnetically consistent model for \gls{ris} with arbitrarily spaced unit cells, which are modeled as loaded wire dipoles. This model is based on the concept of mutually coupled antennas and accounts for near- and far-field channel conditions, thus offering great analytical tractability while ensuring electromagnetic consistency. Indeed, by capitalizing on the DDA, the polarizabilities of general natural or engineered materials (as is the case for \glspl{ris}) are modeled by adjusting the parameters of the wire dipoles and the loads~\cite{504309, faqiri2022physfad, YURKIN2007558}. 

In~\cite{EuCAP_2023}, the authors have introduced a closed-form analytical formulation of the model in~\cite{Gra21}, while in~\cite{Qian20} the model is exploited to derive the optimal \gls{ris} configuration that maximizes the received power in a single-user and single-antenna setup. A closed-form expression is given in the absence of mutual coupling, whereas a convergent iterative algorithm is proposed in the presence of mutual coupling. Such optimization framework is generalized in~\cite{Abrardo2021} for the multi-\gls{ris} multi-user \gls{mimo} interference channel. 

With the exception of~\cite{Abrardo2021}, such works are applicable to free-space propagation in the presence of an RIS. On the other hand, the authors of~\cite{Abrardo2021} model the multipath from \glspl{eso} by relying on a statistical fading model, which is superimposed (additive component) to the RIS-induced scattering. Inherently, this approach does not account for the interactions (e.g., multiple reflections) between the RIS and the scattering objects. Nevertheless, recent results show that multiple reflections between closely-located scatterers (engineered or not) cannot be ignored \cite{9852985}.

{\bf Contributions.} In this letter, we further extend the model in~\cite{Gra21} to account for the multipath propagation originated from the presence of \glspl{eso}, which are modeled as a collection of loaded wire dipoles according to the DDA. In particular, the parameters of the dipoles modeling the \glspl{eso} (e.g., the length and the loads) are chosen to match the material properties of the objects~\cite{504309}, while the RIS loads are tunable on-demand. The proposed approach inherently captures the interactions between the RIS and the \glspl{eso} unlike the additive model in~\cite{Abrardo2021}, which we prove is obtained as a particular case.

Based on the proposed model, we formulate a provably convergent optimization algorithm for computing the optimal load impedances of the RIS and the transmit precoding that maximize the \gls{smse} in a multiple-input single-output multi-user network, dubbed as \emph{\name{}}. Compared with the state-of-the-art optimization algorithm introduced in~\cite{Abrardo2021}, \name{} is shown to provide a higher sum-rate while requiring fewer iterations and less execution time.}

\textbf{Notation}. Matrices and vectors are denoted in bold font. $(\cdot)^{\tran}$, $(\cdot)^{\herm}$, and $\tr(\cdot)$ stand for transposition, Hermitian transposition, and trace of a square matrix, respectively. $\|\cdot\|$ denotes the spectral norm and $\|\cdot \|_{\mathrm{F}}$ the Frobenius norm of a matrix. $\I_N$ denotes the identity matrix of size $N$ and $\0_{N \times M}$ the all-zero $N\times M$ matrix, \change{Lastly, $j=\sqrt{-1}$ is the imaginary number.}

\section{System model}

We consider a network comprising a transmitter with $M$ antennas, $L$ single-antenna receivers, and an \gls{ris} equipped with $N$ reconfigurable (through tunable impedances) unit cells. Also, $N_s$ \glspl{eso} are available in the considered propagation environment. According to the DDA for natural (\glspl{eso}) or engineered (RIS)~\cite{YURKIN2007558, 504309} scatterers, the $N$ unit cells and $N_s$ \glspl{eso} are modeled as loaded wire dipoles. The loads of the \glspl{eso} are not tunable, while the loads of the RIS are tunable on-demand for enhancing the communication performance. This allows us to adopt a unified model for RIS and \glspl{eso}.

Based on the model in~\cite{Gra21}, the end-to-end channel that accounts for the RIS and \glspl{eso} is formulated as~\cite[Eq. (6)]{Abrardo2021}
\ifsingle
\else
\begin{align}\label{eq:H_e2e}
    \H_{\rme} & =  \left(\I_{L} + \Z_{\rmR\rmR}\Z_{\rmL}^{-1}\right)^{-1}\Big[\Z_{\rmR\rmT}-\Z_{\rmR\rmE}\big(\Z_{\rmE\rmE}+\Z_{\rmscat}\big)^{-1}\Z_{\rmE\rmT}\Big]\nonumber \\
    &\phantom{=}\times \left(\Z_{\rmT\rmT}+\Z_{\rmG}\right)^{-1} \in \Compl^{L\times M} 
\end{align} 
\fi
where $\{\mathrm{T,R,E}\}$ denote the transmitter, the receivers, and the scattering environment that includes the RIS and \glspl{eso}. Compared with~\cite{Gra21} and~\cite{Abrardo2021}, the contribution of the \glspl{eso} is explicitly taken into account, as detailed next. Specifically, $\Z_{\rmG} \in \Compl^{M\times M}$ and $\Z_{\rmL}  \in \Compl^{L\times L}$ are the diagonal matrices containing the internal impedances of the voltage generators at the transmitter and the load impedances at the receivers, respectively; $\Z_{\rmT\rmT} \in \Compl^{M\times M}$ and $\Z_{\rmR\rmR} \in \Compl^{L\times L}$ are the matrices containing the self and mutual impedances at the transmitter and receivers, respectively; and $\Z_{\rmR\rmT}\in \Compl^{L\times M}$ is the channel matrix of the direct link between the transmitter and receivers.

The term $\Z_{\rmR\rmT}-\Z_{\rmR\rmE}\big(\Z_{\rmE\rmE}+\Z_{\rmscat}\big)^{-1}\Z_{\rmE\rmT}$ accounts for the signal scattered by the RIS and \glspl{eso}. Specifically, $\Z_{\rmE\rmT}\in \Compl^{(N+N_s)\times M}$ and $\Z_{\rmR\rmE}\in \Compl^{L\times (N+N_s)}$ are the channel matrices containing the mutual impedance between the transmitter and the scattering environment, and between the scattering environment and the receivers, respectively. Also, $\Z_{\rmE\rmE}\in\Compl^{(N+N_s)\times (N+N_s)}$ is the matrix containing the self and mutual impedances between the RIS and the \glspl{eso}, including the mutual coupling among the unit cells of the RIS. For further analysis, it is convenient to separate the contribution from the RIS and the \glspl{eso}. By denoting with the subscripts $\rm S$ and $\rm O$ the contributions of the RIS and the \glspl{eso}, respectively, the matrix $\Z_{\rmE\rmE}$ can be partitioned into four blocks, as follows:
\begin{align}
    \Z_{\rmE\rmE} & = \begin{bmatrix}
    \Z_{\rmO\rmO} & \Z_{\rmO\rmS} \\
    \Z_{\rmS\rmO} & \Z_{\rmS\rmS}
    \end{bmatrix}\in \Compl^{(N+N_s)\times (N+N_s)}.
\end{align}

By using a similar notation, the matrices $\Z_{\rmE\rmT}$ and $\Z_{\rmR\rmE}$ can be partitioned as follows:
\begin{align}
    \Z_{\rmR\rmE} = \begin{bmatrix}\Z_{\rmR\rmO} & \Z_{\rmR\rmS}\end{bmatrix}, \quad
    \Z_{\rmE\rmT} = \begin{bmatrix}\Z_{\rmO\rmT}^\tran & \Z_{\rmS\rmT}^\tran\end{bmatrix}^\tran.
\end{align}

Lastly, the diagonal matrix $\Z_{\rmscat}$ contains the tunable load impedances of the RIS and the load impedances of the \glspl{eso} that model their material properties. Similar to $\Z_{\rmE\rmE}$, $\Z_{\rmE\rmT}$, and $\Z_{\rmR\rmE}$, the matrix $\Z_{\rmscat}$ can be partitioned as follows:
\begin{align}
    \Z_{\rmscat} &= \begin{bmatrix}
    \Z_{\rmUS} & \change{\0_{N_s \times N}} \\
    \change{\0_{N \times N_s}} & \Z_{\rmRIS}
    \end{bmatrix}\in \Compl^{(N+N_s)\times (N+N_s)}
\end{align}
where $\Z_{\rmUS}\in\Compl^{N_s\times N_s}$ is the diagonal matrix containing the loads of the \glspl{eso} and $\Z_{\rmRIS} \in \Compl^{N\times N}$ is the diagonal matrix containing the tunable impedances of the unit cells of the \gls{ris}. Each entry of the matrices of self and mutual impedances in~\eqref{eq:H_e2e} can be computed via the analytical frameworks in~\cite{EuCAP_2023, Phan18}, while the matrix $\Z_{\rmUS}$ is determined by the material properties, e.g., the permittivity, of the \glspl{eso}, and is thus assumed to be fixed and given. On the other hand, the matrix $\Z_{\rmRIS}$ can be optimized, and, assuming that the RIS is nearly passive, is modeled as follows:
\begin{align} \label{eq:Zris}
    \Z_{\rmRIS} = \mathrm{diag}\left[\mathrm{R}_0 + j x_n\right]_{n=1}^N
\end{align}
where $\mathrm{R}_0 \ge 0$ is the resistance of each load, which accounts for the internal losses of the tuning circuits. If $\mathrm{R}_0 = 0$, the RIS is lossless. The constraint $\mathrm{R}_0 \ge 0$ ensures that the RIS does not amplify the incident signal. It is customary to assume that $\mathrm{R}_0$ is fixed, since it depends on the technology being used and the best choice is $\mathrm{R}_0 = 0$ to minimize the losses. On the other hand, the reactance $x_i\in \mathcal{Q}$ is tunable, with $\mathcal{Q}$ denoting the set (continuous or discrete) of feasible values for it. 

To simplify the writing, we use the notation $\Z_{\rmR\rmL} \triangleq (\I_L+\Z_{\rmR\rmR} \Z_{\rmL}^{-1})^{-1}\in\Compl^{L\times L}$ and $\Z_{\rmT\rmG} \triangleq \left(\Z_{\rmT\rmT}+\Z_{\rmG}\right)^{-1}\in\Compl^{M\times M}$. Hence, the received signal at the $\ell$-th \gls{ue}\change{, with $\ell=1,\ldots,L$,} is given by
\ifsingle
\else
\begin{align}\label{eq:y_l}
    y_{\ell} & = \h_{\rme,\rmd,\ell}\W\s+ \h_{\rme,\ell}(\Z_{\rmris})\W\s + n_{\ell}\\
    & = \big(\h_{\rme,\rmd,\ell} + \h_{\rme,\ell}(\Z_{\rmris})\big)\w_{\ell}s_{\ell} \nonumber \\
    & \phantom{=}+ \sum\nolimits_{k=1, k\neq \ell}^L \big(\h_{\rme,\rmd,\ell} + \h_{\rme,\ell}(\Z_{\rmris})\big)\w_{k}s_{k} + n_{\ell} \nonumber
\end{align}
\fi
where $\W\in\Compl^{M\times L}$ denotes the precoding matrix at the transmitter, with $\|\W\|_{\mathrm{F}}^2\leq P$ and $P$ being the power budget; $\s = [s_1,\ldots,s_L]^\tran \in \Compl^{L\times 1}$ is the transmit data vector, with $\Exp[\|\s\|^2]=1$; and $\n\sim\mathcal{CN}(\0,\sigma_n^2\I_L)$ is the noise vector at the receivers. Also, the following equivalent channels for the $\ell$-th \gls{ue} are introduced:
\begin{align} 
    \h_{\rme,\rmd,\ell} &\triangleq \z_{\rmR\rmL,\ell}\Z_{\rmR\rmT}\Z_{\rmT\rmG}\in\Compl^{1\times M} \label{Eq:UEchannels}\\
    \h_{\rme,\ell}(\Z_{\rmris}) & \triangleq  -  \z_{\rmR\rmL,\ell}\Z_{\rmR\rmE}\big(\Z_{\rmE\rmE}+\Z_{\rmscat}\big)^{-1}\Z_{\rmE\rmT}\Z_{\rmT\rmG}\in\Compl^{1\times M}\nonumber
\end{align}
with the lower-case bold letters denoting the $\ell$-th row of the corresponding matrices in~\eqref{eq:H_e2e}. With this notation, the sum-rate is given as follows:\change{
\begin{align} \label{Eq:SumRate}
\gamma_{\ell} & = \frac{|[\h_{\rme,\rmd,\ell} \!+\! \h_{\rme,\ell}(\Z_{\rmris})]\w_{\ell}|^2}{\displaystyle \sum\nolimits_{k=1, k\neq \ell}^L \big\rvert[\h_{\rme,\rmd,\ell} \!+\! \h_{\rme,\ell}(\Z_{\rmris})]\w_{k}\big\rvert^2\!\!\!+\!\sigma_n^2} \nonumber \\
    R &  = \sum\nolimits_{\ell=1}^L\log_2\left(1+\gamma_{\ell}\right).
\end{align}}
To simplify the notation and for further analysis, we introduce the following impedance matrices:
\begin{align}
    \overline{\Z}_{\rmO\rmO} & \triangleq \Z_{\rmO\rmO}+\Z_{\rmUS} \in \Compl^{N_s\times N_s}\\
    \Z_{\rmR\rmO\rmT} & \triangleq \Z_{\rmR\rmT} - \Z_{\rmR\rmO}\overline{\Z}_{\rmO\rmO}^{-1}\Z_{\rmO\rmT} \in \Compl^{L\times M}\label{eq:z_rot} \\
    \Z_{\rmR\rmO\rmS} & \triangleq \Z_{\rmR\rmO}\overline{\Z}_{\rmO\rmO}^{-1}\Z_{\rmO\rmS}-\Z_{\rmR\rmS} \in \Compl^{L\times N}\\
    \Z_{\rmS\rmO\rmS} & \triangleq  -\Z_{\rmS\rmO}\overline{\Z}_{\rmO\rmO}^{-1}\Z_{\rmO\rmS} \in \Compl^{N\times N}\\
    \Z_{\rmS\rmO\rmT} & \triangleq \Z_{\rmS\rmO}\overline{\Z}_{\rmO\rmO}^{-1}\Z_{\rmO\rmT}-\Z_{\rmS\rmT}\in \Compl^{N\times M}.
\end{align}

As remarked in~\cite{Qian20} and~\cite{Abrardo2021}, the main difficulty when optimizing the sum-rate in~\eqref{Eq:SumRate} is given by the computation of the inverse matrix $\big(\Z_{\rmE\rmE}+\Z_{\rmscat}\big)^{\!-1}$, since it is a non-diagonal matrix in the presence of mutual coupling. To deal with this matrix inversion, we invoke the Schur complement applied to block matrices~\cite{boyd_vandenberghe_2004}. Specifically, assuming that $\overline{\Z}_{\rmO\rmO}$ is invertible, the matrix $\big(\Z_{\rmE\rmE}+\Z_{\rmscat}\big)^{-1}$ can be rewritten as shown in~\eqref{eq:Schur_compl} at the top of this page. Thus,~\eqref{eq:H_e2e} and~\eqref{Eq:UEchannels} can be simplified, respectively, as follows:
\begin{align}
    & \H_{\rme}  \!=\! \Z_{\rmR\rmL}\Big[\Z_{\rmR\rmO\rmT} - \Z_{\rmR\rmO\rmS}\big(\Z_{\rmS\rmS}+\Z_{\rmS\rmO\rmS}+\Z_{\rmris}\big)^{-1}\Z_{\rmS\rmO\rmT}\Big]\Z_{\rmT\rmG}, \nonumber\\
     &\h_{\rme,d,\ell}  = \z_{\rmR\rmL,\ell}\Z_{\rmR\rmO\rmT}\Z_{\rmT\rmG},\label{eq:H_e2e_3}\\
   & \h_{\rme,\ell}(\Z_{\rmris})  = - \z_{\rmR\rmL,\ell}\Z_{\rmR\rmO\rmS}\big(\Z_{\rmS\rmS}+\Z_{\rmS\rmO\rmS}+\Z_{\rmris}\big)^{-1}\Z_{\rmS\rmO\rmT}\Z_{\rmT\rmG}. \nonumber
\end{align}

The analytical expression of the end-to-end channel $\H_{\rme}$ in \eqref{eq:H_e2e_3} is conveniently formulated, since the optimization variables, i.e., the diagonal matrix $\Z_{\rmris}$, is decoupled from the non-diagonal matrices $\Z_{\rmS\rmS}$ and $\Z_{\rmS\rmO\rmS}$ that account for the mutual coupling between the unit cells of the RIS and for the interactions between the unit cells and the \glspl{eso} (including the mutual coupling among the wire dipoles of the \glspl{eso}), respectively. This facilitates the solution of optimization problems and provides deeper insights into the end-to-end channel.

\ifsingle
\else

{
\begin{figure*}[t!]
\begin{align}\label{eq:Schur_compl}
\footnotesize
    \begin{bmatrix}
    \overline{\Z}_{\rmO\rmO} \!&\! \Z_{\rmO\rmS} \\
    \Z_{\rmS\rmO} \!&\! \Z_{\rmS\rmS}+\Z_{\rmRIS}
    \end{bmatrix} ^{-1} \!\!\!\!= \!\!\begin{bmatrix}
    \overline{\Z}_{\rmO\rmO}^{-1}+\overline{\Z}_{\rmO\rmO}^{-1}\Z_{\rmO\rmS}(\Z_{\rmRIS} + \Z_{\rmS\rmS} - \Z_{\rmS\rmO}\overline{\Z}_{\rmO\rmO}^{-1}\Z_{\rmO\rmS})^{-1}\Z_{\rmS\rmO}\overline{\Z}_{\rmO\rmO}^{-1} \!&\! -\overline{\Z}_{\rmO\rmO}^{-1}\Z_{\rmO\rmS}(\Z_{\rmRIS} + \Z_{\rmS\rmS} - \Z_{\rmS\rmO}\overline{\Z}_{\rmO\rmO}^{-1}\Z_{\rmO\rmS})^{-1} \\
    -(\Z_{\rmRIS} + \Z_{\rmS\rmS} - \Z_{\rmS\rmO}\overline{\Z}_{\rmO\rmO}^{-1}\Z_{\rmO\rmS})^{-1}\Z_{\rmS\rmO}\overline{\Z}_{\rmO\rmO}^{-1} \!&\! (\Z_{\rmRIS} + \Z_{\rmS\rmS} - \Z_{\rmS\rmO}\overline{\Z}_{\rmO\rmO}^{-1}\Z_{\rmO\rmS})^{-1}
    \end{bmatrix}
\end{align}
\hrulefill \vspace{-0.5cm}
\end{figure*}}
\fi

\subsection{Model Novelty: Modeling the \glspl{eso}}\label{subsec:modelnovelty}
The analytical expression of $\H_{\rme}$ in \eqref{eq:H_e2e_3} clearly unveils the difference between the proposed model for the \glspl{eso} (i.e., the multipath) and closely related papers. Specifically, the authors of~\cite{Abrardo2021} have considered an additive statistical model for the multipath based on conventional fading models. The model in~\cite{Abrardo2021} can be retrieved \change{as a special case from our proposed} $\H_{\rme}$ in \eqref{eq:H_e2e_3} by setting $\Z_{\rmS\rmO}=\Z_{\rmO\rmS}^\tran=\0_{N \times N_s}$, which yields $\Z_{\rmR\rmO\rmS} = -\Z_{\rmR\rmS}$, $\Z_{\rmS\rmO\rmS}=\0_{N \times N}$, and $\Z_{\rmS\rmO\rmT}=-\Z_{\rmS\rmT}$. As a result, $\H_{\rme}$ in \eqref{eq:H_e2e_3} simplifies to
\begin{align} \label{eq:NoInteractions}
    &\H_{\rme} = \Z_{\rmR\rmL}\Big[\Z_{\rmR\rmO\rmT} - \Z_{\rmR\rmO\rmS}\big(\Z_{\rmS\rmS}+\Z_{\rmris}\big)^{-1}\Z_{\rmS\rmO\rmT}\Big]\Z_{\rmT\rmG}\\
    & = \Z_{\rmR\rmL}\!\Big[\Z_{\rmR\rmT} - \Z_{\rmR\rmO}\overline{\Z}_{\rmO\rmO}^{-1}\Z_{\rmO\rmT}- \Z_{\rmR\rmS}\big(\Z_{\rmS\rmS}+\Z_{\rmris}\big)^{-1}\Z_{\rmS\rmT}\!\Big] \Z_{\rmT\rmG}. \nonumber
\end{align}

The resulting end-to-end channel in~\eqref{eq:NoInteractions} is given by the summation of the free-space channel in~\cite{Gra21} and an additive multipath component given by $ \Z_{\rmR\rmL}\Z_{\rmR\rmO}\overline{\Z}_{\rmO\rmO}^{-1}\Z_{\rmO\rmT}\Z_{\rmT\rmG}$, which accounts for the scattering from the \glspl{eso} and is independent of the RIS, similar to the multipath model considered in~\cite{Abrardo2021}. \change{However, the channel in~\eqref{eq:NoInteractions} inherently ignores the interactions between the \gls{ris} and the \glspl{eso}, thus resulting in less accurate modeling of the physical propragation environment (see, e.g.,~\cite{9852985}) and suboptimal \gls{ris} designs that may fail to fully exploit its properties. Moreover, since the load impedances $\Z_{\rmUS}$ can be arbitrarily chosen to mimic any natural material, while all the other self and mutual impedances depend mainly on the geometry of the scenario (e.g., see~\cite{EuCAP_2023, Phan18}), the proposed model subsumes deterministic and statistical multipath channel models, i.e., in the case where the  locations of the \glspl{eso} are assumed to be unknown (e.g., random).} 

\section{Problem Formulation and Solution}
To optimize $\Z_{\rmris}$ subject to the constraint $\mathrm{R}_0 \ge 0$ in \eqref{eq:Zris} and $\W\in\Compl^{M\times L}$ subject to the constraint $\|\W\|_{\mathrm{F}}^2\leq P$, we consider the \gls{smse} as objective function. The motivation for using the \gls{smse} and its relation with the optimization of the sum-rate in \eqref{Eq:SumRate} is detailed in \cite{Mursia20}. The \gls{smse} is defined as
\ifsingle
\else
\begin{align} \label{eq:smse}
    \mathrm{SMSE} \! & =  \!\sum\nolimits_{\ell=1}^L \sum\nolimits_{k=1}^L |\big(\h_{\rme,\rmd,\ell} + \h_{\rme,\ell}(\Z_{\rmris})\big)\w_{k}|^2 \\
    &-2\!\sum\nolimits_{\ell=1}^L\! \Re\{\!\big(\h_{\rme,\rmd,\ell}\! +\! \h_{\rme,\ell}(\Z_{\rmris})\big)\w_{\ell}\}\!+ \!L(1\!+\!\sigma_n^2)\nonumber.
\end{align}
\fi

Hence, the optimization problem is formulated as follows: 
\ifsingle
\else\begin{align}
    \label{eq:Prob}
        \displaystyle \min_{\W,\,\, \Z_{\rmRIS}} & \displaystyle \sum\nolimits_{\ell,\,\,k=1}^L  |\big(\h_{\rme,\rmd,\ell} + \h_{\rme,\ell}(\Z_{\rmris})\big)\w_{k}|^2\nonumber\\& \displaystyle -2 \sum\nolimits_{\ell=1}^L \Re\{\big(\h_{\rme,\rmd,\ell} + \h_{\rme,\ell}(\Z_{\rmris})\big)\w_{\ell}\} \vspace{0.25cm} \nonumber\\
        \mathrm{s.t.} \quad \!& \|\W\|_{\mathrm{F}}^2\leq P,\,\, \Re\{\Z_{\rmRIS}\}\! =\! \mathrm{R}_0,\,\, \Im\{\Z_{\rmRIS}\}\! \in\! \mathcal{Q}^N\!.
\end{align}
\fi

Due to the presence of the \glspl{eso}, the optimization problem in~\eqref{eq:Prob} is more challenging to solve as compared with those in~\cite{Qian20} and~\cite{Abrardo2021}. In fact, the matrix $\Z_{\rmS\rmO\rmS}$ is, in general, not a diagonal matrix even if the mutual coupling between the \glspl{eso} is negligible, i.e., $\overline{\Z}_{\rmO\rmO}$ is diagonal. This is because the matrices $\Z_{\rmS\rmO}$ and $\Z_{\rmO\rmS}$ are not diagonal matrices, in general. To tackle the problem in~\eqref{eq:Prob}, we decouple it into two convex sub-problems that are solved iteratively, as detailed next.

\subsection{Precoding Optimization}\label{subsubsec:Precoding}
First, we solve the problem in~\eqref{eq:Prob} with respect to the optimization variable $\W$ while keeping $\Z_{\rmRIS}$ fixed. The resulting problem is convex and its solution is \change{found by evaluating the KKT conditions. In this regard, let the Lagrangian and its gradient be
\ifsingle
\else
\begin{align}
    \mathcal{L}(\W,\mu) & = \|\H_{\rme}^\herm(\Z_{\rmRIS}\!)\W\|_\mathrm{F}^2 - 2\tr(\Re\{\H_{\rme}^\herm(\Z_{\rmRIS}\!)\W\}\nonumber \\
    & +\mu(\|\W\|_{\mathrm{F}}^2-P),\nonumber\\
    \nabla \mathcal{L}(\W,\mu) & = \big(\H_{\rme}\H_{\rme}^\herm+\mu\I_M\big)\W - \H_{\rme},\label{eq:Nabla_W}
\end{align}
\fi
respectively. Hence, the optimal closed-form solution is given by letting the expression in~\eqref{eq:Nabla_W} to zero as}~\cite[Eqs. (28), (29)]{Mursia20}
\begin{align}
    \overline{\W}(\Z_{\rmRIS})& \!=\! \bigg(\H_{\rme}^\herm(\Z_{\rmRIS}\!)\H_{\rme}(\Z_{\rmRIS}\!)\!+\!\frac{L\sigma_n^2}{P} \I_M\!\bigg)^{\!-1}\!\!\H_{\rme}^\herm(\Z_{\rmRIS}\!)\nonumber \\
    \W(\Z_{\rmRIS}) & \!=\! \sqrt{P}\,\overline{\W}(\Z_{\rmRIS})/(\|\overline{\W}(\Z_{\rmRIS})\|_{\mathrm{F}}) \in \Compl^{M\times L}\label{eq:precoder}.
\end{align}

\setlength{\textfloatsep}{5pt}
\begin{algorithm}[t!]
\footnotesize
  \caption{\change{\name}}\label{alg:A1}
  \change{\textbf{Input:} Threshold $\epsilon$, channel $\H_{\rme}$, power budget $P$, noise power $\sigma_n^2$
  \begin{algorithmic}[1]
     \State Initialize $i=1$, $\Z_{\rmris}^0 = \0$, $\Z_{\rmris}^1$; $\mathrm{SMSE}(\W,\,\Z_{\rmris}^{i})$ defined in~\eqref{eq:smse}
     \While {$|\mathrm{SMSE}(\W,\,\Z_{\rmris}^{i})-\mathrm{SMSE}(\W,\,\Z_{\rmris}^{i-1})|>\epsilon$}
        \State Update $\W$ as in~\eqref{eq:precoder}
        \State Update $\deltab$ as in~\eqref{eq:delta_b}, $\Deltab=\mathrm{diag}(\deltab^\herm)$
        \State $\Z_{\rmRIS}^{i+1} = \Z_{\rmRIS}^{i} + j \Im\{\Deltab\}$
        \State Projection of the imaginary part of $\Z_{\rmRIS}^{i+1}$ onto $\mathcal{Q}^N$
        \State $i=i+1$
     \EndWhile
  \end{algorithmic} 
  \textbf{Output:} $\Z_{\rmRIS}^{i}$, $\W$}
\end{algorithm}

\subsection{\gls{ris} Optimization}\label{subsubsec:iterative_Neuman}
Then, we solve the problem in \eqref{eq:Prob} with respect to the optimization variable $\Z_{\rmRIS}$ while keeping $\W$ fixed. To this end, we devise an iterative algorithm that, similar to \cite{Qian20} and \cite{Abrardo2021}, leverages the repeated application of the Neumann series for efficiently tackling the inversion of $\big(\Z_{\rmS\rmS}+\Z_{\rmS\rmO\rmS}+\Z_{\rmris}\big)$. With respect to \cite{Qian20} and \cite{Abrardo2021}, however, we introduce an improved algorithm that accounts for the necessary conditions to make the Neumann series accurate by design. This is detailed next.

Specifically, at the $(i+1)$th iteration, the optimization variable $\Z_{\rmRIS}$ is set to $\Z_{\rmRIS}^{i+1} = \Z_{\rmRIS}^{i} + \Deltab$, where $\Deltab=\mathrm{diag}[\deltab^\herm]\in\Compl^{N\times N}$ is a diagonal matrix containing small improvements to the \gls{ris} configuration. With this approximation, we obtain \change{
\ifsingle
\else
\begin{align}
     & \h_{\rme,\ell}^{i+1}(\Z_{\rmris})  = -\z_{\rmR\rmL,\ell}\Z_{\rmR\rmO\rmS}\big(\Z_{\rmS\rmS}+\Z_{\rmS\rmO\rmS}+\Z_{\rmris}^{i+1}\big)^{-1}\Z_{\rmS\rmO\rmT}\Z_{\rmT\rmG} \nonumber\\
    & \approx - \z_{\rmR\rmL,\ell}\Z_{\rmR\rmO\rmS}\Big[\G^i -\G^i\Deltab\G^i \Big]\Z_{\rmS\rmO\rmT}\Z_{\rmT\rmG}\nonumber\\
    & =\!\bar{\deltab}^\tran\big[\mathrm{diag}(\z_{\rmR\rmL,\ell}\Z_{\rmR\rmO\rmS}\G^i)\G^i \!-\! \z_{\rmR\rmL,\ell}\Z_{\rmR\rmO\rmS}\G^i \big]\Z_{\rmS\rmO\rmT}\Z_{\rmT\rmG}\label{eq:h_neumann_l}
\end{align}
\fi
where $\G^i \triangleq (\Z_{\rmS\rmS}+\Z_{\rmS\rmO\rmS}+\Z_{\rmris}^{i})^{-1}$ and $\bar{\deltab} \triangleq \begin{bmatrix}
    \deltab^\tran & 1
    \end{bmatrix}^\tran$. Therefore, we have
\begin{align}
    \big(\h_{\rme,\rmd,\ell} + \h_{\rme,\ell}^{i+1}(\Z_{\rmris})\big)\w_{j} \approx \bar{\deltab}^\herm\bar{\H}^{i+1}_{\ell}\w_j
\end{align}
with
\begin{align}
    \bar{\H}^{i+1}_{\ell} & \triangleq \begin{bmatrix}
    \mathrm{diag}(\z_{\rmR\rmL,\ell} \Z_{\rmR\rmO\rmS}\G^i)\G^i\Z_{\rmS\rmO\rmT}\Z_{\rmT\rmG} \\
    \h_{\rme,\rmd,\ell} - \z_{\rmR\rmL,\ell}\Z_{\rmR\rmO\rmS}\G^i\Z_{\rmS\rmO\rmT}\Z_{\rmT\rmG}
    \end{bmatrix}\\
    & =  \begin{bmatrix} (\bar{\H}^{i+1}_{\rmR, \ell})^\tran & (\bar{\h}^{i+1}_{\rmd, \ell})^\tran \end{bmatrix}^\tran \in\Compl^{N+1\times M}.
\end{align}
}\indent As a result, the \gls{smse} can be approximated as follows:
\begin{align}
    \overbar{\mathrm{SMSE}} & \approx \displaystyle \sum_{\ell,k=1}^L  |\bar{\deltab}^\herm \bar{\H}^{i+1}_\ell\w_{k}|^2 -2\sum_{\ell=1}^L \Re\{\bar{\deltab}^\herm \bar{\H}^{i+1}_\ell\w_{\ell}\} 
    \label{eq:smse_b}
\end{align}
\indent The approximation in~\eqref{eq:h_neumann_l} is accurate provided that the condition $\|\Deltab\G^i\|\ll 1$ is fulfilled at each iteration. In \cite{Qian20} and \cite{Abrardo2021}, this condition is taken into account by choosing very small values for the elements of $\deltab$. This approach, however, ignores the actual values of $\G^i$. To ensure that the Neumann series approximation is accurate by design at each iteration of the algorithm, we include the condition $\|\Deltab\G^i\|\ll 1$ as a constraint of the optimization problem to be solved. This leads to the following simplified and robust problem formulation:
\begin{align}
    \label{eq:Prob2}
        \displaystyle \min_{\bar{\deltab}} & \displaystyle  \sum\nolimits_{\ell,k=1}^L |\bar{\deltab}^\herm \bar{\H}^{i+1}_\ell\w_{k}|^2 -2\sum\nolimits_{\ell=1}^L \Re\{\bar{\deltab}^\herm \bar{\H}^{i+1}_\ell\w_{\ell}\} \nonumber\\
        \mathrm{s.t.} \quad & \bar{\delta}_{N+1} = 1, \quad |{\delta}_n|\leq{1}/{\|\G^{i}\|},\, n=1,\ldots,N
\end{align}
where the constraint on $|{\delta}_n|$ ensures that the Neumann series approximation is sufficiently accurate at each iteration. \change{As in \eqref{eq:Nabla_W}--\eqref{eq:precoder}, the convex optimization problem in~\eqref{eq:Prob2} is solved by setting the gradient of the Lagrangian to zero as
\ifsingle
\else
\begin{align}
    \mathcal{L}(\bar{\deltab},\mub)& = \sum\nolimits_{\ell=1}^L\big( \|\bar{\deltab}^\tran \bar{\H}_{\ell}^{i+1}\W\|^2-2\Re\{\bar{\deltab}^\tran \bar{\H}_{\ell}^{i+1}\w_{\ell}\}\big)\nonumber \\
    &+\sum\nolimits_{n=1}^N \mu_n(|\delta_n|^2-1/\|\G^i\|)\nonumber \\
    \nabla\mathcal{L}(\bar{\deltab},\mub)& =\sum\nolimits_{\ell=1}^L\big(\bar{\H}_{\ell}^{i+1}\W\W^\herm(\bar{\H}_{\ell}^{i+1})^\herm\bar{\deltab}-\bar{\H}_{\ell}^{i+1}\w_{\ell}\big)\nonumber\\
    & + \mathrm{diag}(\mub)\bar{\deltab},
\end{align}
\fi
}which has the following closed-form solution~\cite[Eq. (22)]{Mursia20}:
\begin{align}
    \b & = \sum\nolimits_{\ell=1}^L \big( \bar{\H}^{i+1}_{\rmR, \ell}\w_{\ell}-\bar{\H}^{i+1}_{\rmR, \ell}\W\W^\herm(\bar{\h}^{i+1}_{\rmd, \ell})^\herm\big)\\
    \tilde{\deltab} & = \Big(\sum\nolimits_{\ell=1}^L\bar{\H}^{i+1}_{\rmR, \ell}\W\W^\herm(\bar{\H}^{i+1}_{\rmR, \ell})^\herm+\sigma_n^{2}\I_N\Big)^{-1}\b \\
    \deltab & = {\tilde{\deltab}}/\big({\max\,\{|\tilde{\delta}_n|\}_{n=1}^N}\,  {\|\G^i\|}\big)\in \Compl^{N\times 1}\label{eq:delta_b}.
\end{align}
The normalization in~\eqref{eq:delta_b} aims to strike a trade-off between the accuracy of the Neumann series approximation and the speed of convergence of the algorithm, i.e., the optimum is obtained when the inequality constraint in \eqref{eq:Prob2} is an equality. Hence, the small improvements $\boldsymbol{\delta}$ are chosen as large as possible depending on the matrix $\G^i$ at each iteration.

The complete algorithm that iterates between the solutions in~\eqref{eq:precoder} and~\eqref{eq:delta_b}\change{, dubbed as \emph{\name},} is provided in Algorithm~\ref{alg:A1}. Specifically, once~\eqref{eq:delta_b} is computed, the value of $\Z_{\rmRIS}$ is updated as $\Z_{\rmRIS}^{i+1} = \Z_{\rmRIS}^{i} + j \Im\{\Deltab\}$, with $\Deltab = \mathrm{diag}(\deltab^\herm)$, in order to preserve the constraint on the real part of $\Z_{\rmRIS}$. Also, the imaginary part of $\Z_{\rmRIS}^{i+1}$ is projected onto the feasible set $\mathcal{Q}^N$. Compared with existing solutions, \change{\name} benefits from the inherent low-complexity of Algorithm~\ref{alg:A1}, since the latter iterates between two simple closed-form expressions.
\ifsingle
\else
\begin{figure}[t!]
    \centering
    \includegraphics[width=0.53\columnwidth]{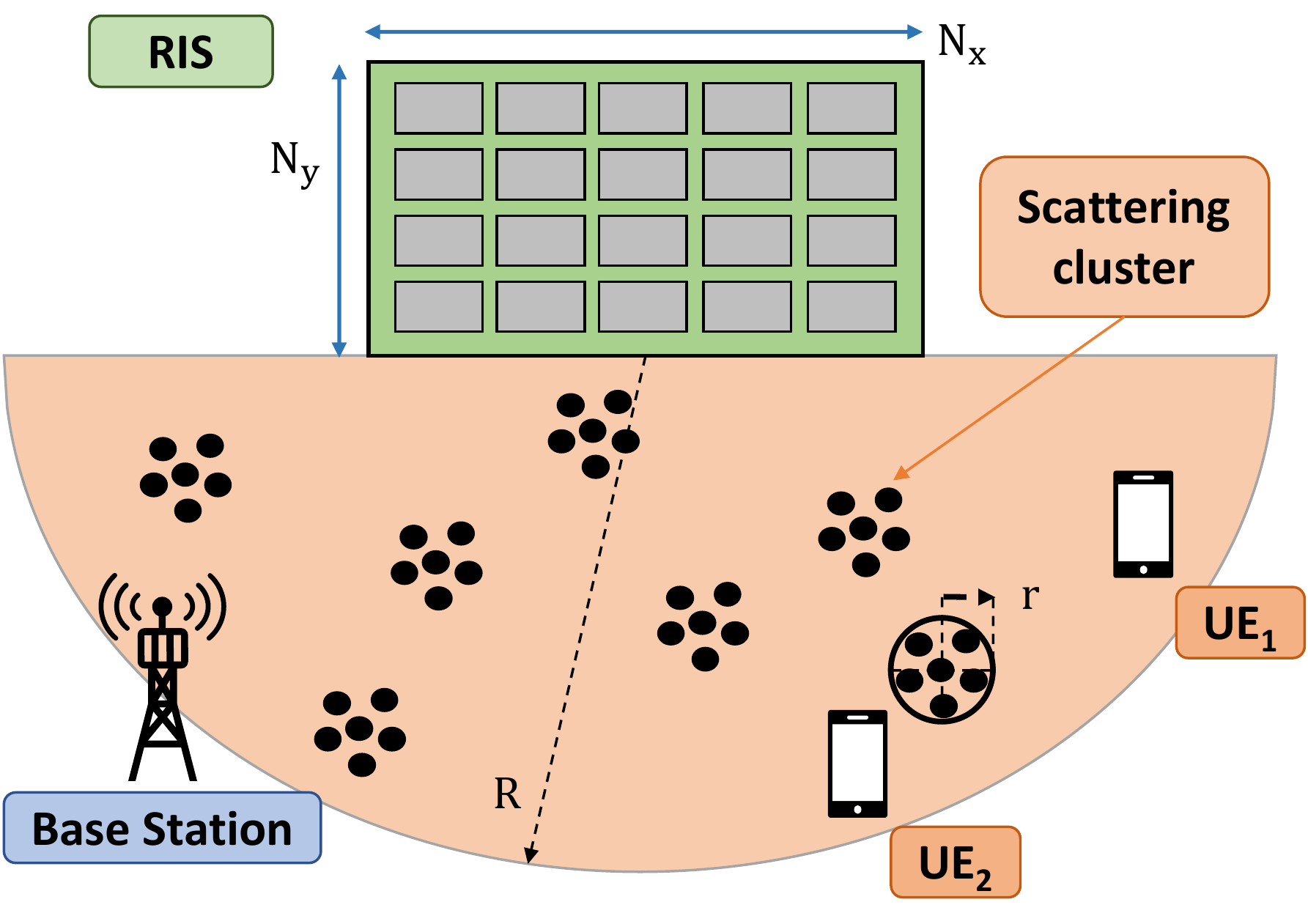}
    \caption{Considered deployment scenario (top view).}
    \label{fig:scenario_UNOT} 
\end{figure}
\fi

\subsection{Computational Complexity and Convergence Analysis}
The computational complexity of~\eqref{eq:precoder} and~\eqref{eq:delta_b} is $\mathcal{O}(M^2(M+3L))$ and $\mathcal{O}(N(3N^2+N(L+1)+3LM+M+L))$, respectively. The complexity of~\eqref{eq:delta_b} is dominated by the inversion of an $N\times N$ matrix and is greater than the complexity of~\eqref{eq:precoder}, since usually $N \gg M$. Therefore, the total complexity of Algorithm~\ref{alg:A1} is $\mathcal{O}(N(3N^2+N(L+1)+3LM+M+L))$ multiplied by the number of required iterations to converge.

The proposed alternating optimization algorithm decouples the original non-convex problem in~\eqref{eq:Prob} into a series of convex sub-problems in the two optimization variables taken separately. \change{Hence, the global objective function, i.e., the \gls{smse}, is a non-increasing function across two consecutive iterations. Moreover, since it is, by definition, lower-bounded by zero, Algorithm~\ref{alg:A1} converges to a critical point of the problem in \eqref{eq:Prob}, which is a sufficient but not necessary condition for optimality~\cite{Mursia20, Abrardo2021}.}
\ifsingle
\else
\begin{figure}
    \begin{tabularx}{\linewidth}{XXX}
       \includegraphics[width=0.454\columnwidth]{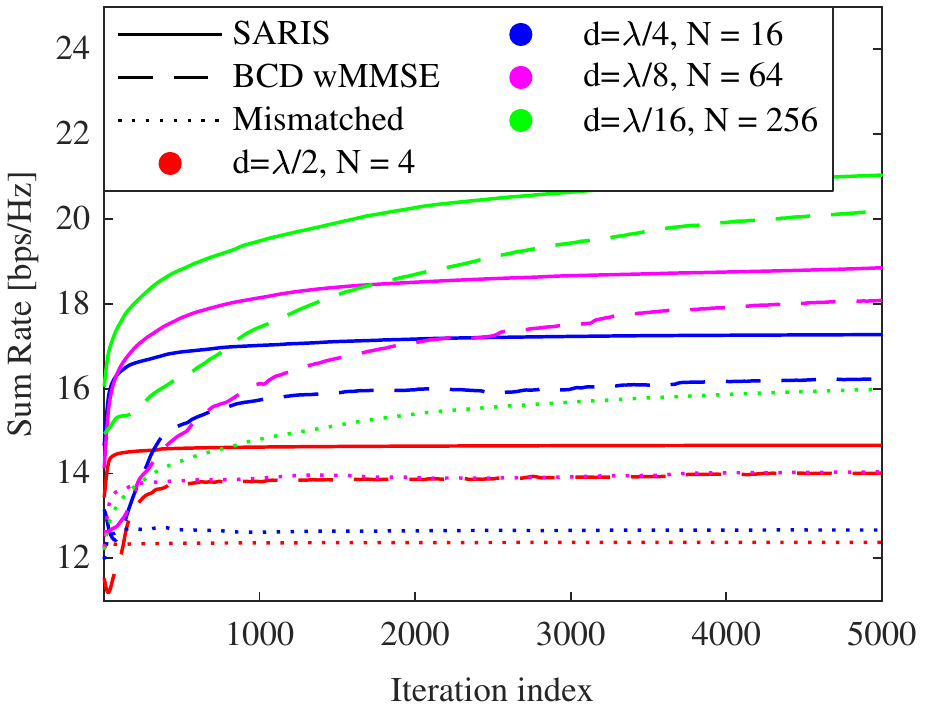}
       \vspace{-4mm}
   \caption{Convergence of Algo.~\ref{alg:A1} vs. $d$.}\label{fig:sumrate_N}%

   &
        
       \includegraphics[width=0.446\columnwidth]{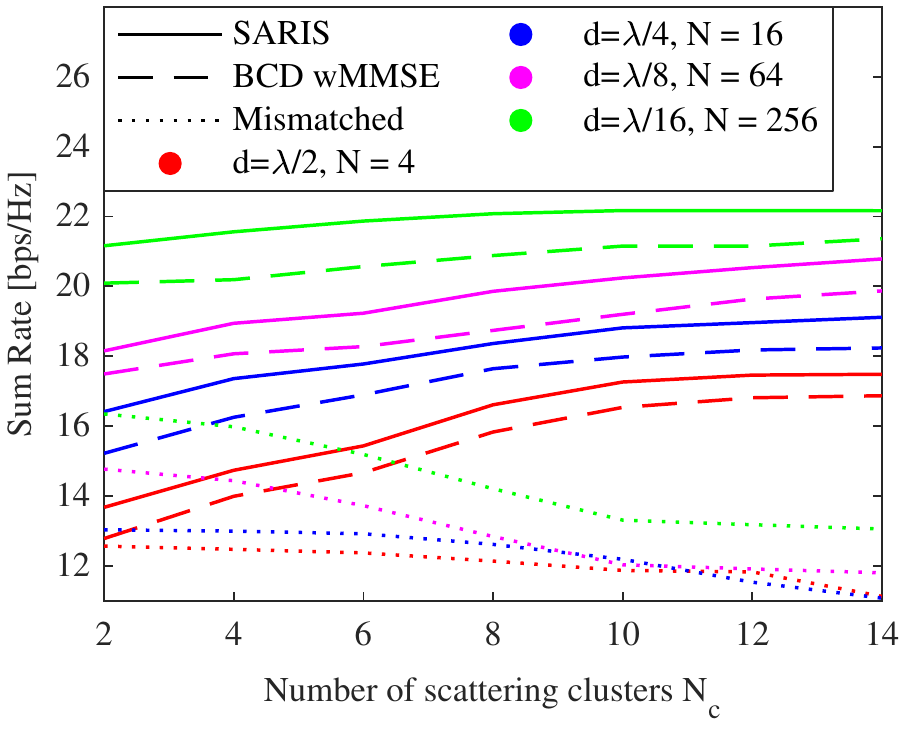}
       \vspace{-4mm}
   \caption{Average sum-rate vs. $N_c$.} \label{fig:N_c}
       \end{tabularx}
       \vspace{-7mm}
\end{figure}
\fi

\ifsingle
\else
\begin{table}[t!]
\caption{Simulation parameters.}
\label{tab:params}
\centering
\resizebox{1\linewidth}{!}{%
\renewcommand{\arraystretch}{1.0}
\begin{tabular}{|c|c|c|c|c|c|}
\hline
\cellcolor[HTML]{EFEFEF} \textbf{Parameter} & \textbf{Value} &\cellcolor[HTML]{EFEFEF} \textbf{Parameter} & \textbf{Value}&\textbf{Parameter} & \textbf{Value} \\
\hline
 \cellcolor[HTML]{EFEFEF}  $N_c$  & $4$ & \cellcolor[HTML]{EFEFEF} $\p_{\rmue2}$ &  $[20,\, 24]\lambda$ &\cellcolor[HTML]{EFEFEF} $\lambda$  & $6$~cm\\
\hline 
 \cellcolor[HTML]{EFEFEF} $\mathrm{R}$ & $40\lambda$ &\cellcolor[HTML]{EFEFEF} $\mathrm{r}$ & $\lambda$  &
\cellcolor[HTML]{EFEFEF} $N_{\rmO}$ & $50$\\
\hline 
\cellcolor[HTML]{EFEFEF} $\rm{R}_0$ & $0.2~\Omega$  &
\cellcolor[HTML]{EFEFEF} $M$ & $4$ & \cellcolor[HTML]{EFEFEF} $\mathcal{Q}$ & $[-302.50, -19.66] \text{ }\Omega$ \\
\hline
\cellcolor[HTML]{EFEFEF} $\p_{\rmue1}$ & $[16,\, 24]\lambda$  &
\cellcolor[HTML]{EFEFEF} $\p_{\rmbs}$ & $[0,\, 0]\lambda$ & \cellcolor[HTML]{EFEFEF} $\p_{\rmris}$ & $[0,\, 40]\lambda$ \\
\hline
\cellcolor[HTML]{EFEFEF} $\Z_{\rmG}$ & $50\,\I_M~\Omega$  &
\cellcolor[HTML]{EFEFEF} $\Z_{\rmL}$ & $50\,\I_L~\Omega$ & \cellcolor[HTML]{EFEFEF} $\Z_{\rmUS}$ & $\0_{N_s \times N_s}$ \\
\hline
\end{tabular}%
}
\renewcommand{\arraystretch}{1}
\end{table}
\fi

\section{Numerical Results} \label{sec:NumRes}
To evaluate the performance of Algorithm~\ref{alg:A1}, we consider the scenario sketched in Fig.~\ref{fig:scenario_UNOT}, where \change{$L=2$} UEs are present in the area of interest\change{, unless otherwise stated.} The simulation parameters are listed in Table~\ref{tab:params}\change{, where we remark that $\Z_{\rmUS} = \0$ models lossless metallic scattering objects.} \change{The inter-distance between the unit cells of the RIS is equal to $d$ along both the $x$- and $y$-axis.} The number of unit cells along the $x$- and $y$-axis is $N_x = N_y = \sqrt{N}$, respectively. The locations of the transmitter, UEs, and the midpoint of the RIS are denoted by $\p_{\rmbs}$, $\p_{\rmue1}$, $\p_{\rmue2}$, and $\p_{\rmris}$, respectively. \change{The $N_s$ \glspl{eso} are distributed in $N_c$ randomly located scattering clusters within a semicircle of radius $\mathrm{R}$ from the \gls{ris}.} Each cluster comprises $N_{\rmO}$ loaded wire dipoles, which are uniformly distributed (at random) within a disk of radius $\mathrm{r}$ centered at the locations of the clusters. The length of all wire dipoles (RIS and \glspl{eso}) is $\lambda/2$, where $\lambda$ is the wavelength. All wire dipoles (RIS and \glspl{eso}) are aligned along the $z$-axis, i.e., Fig.~\ref{fig:scenario_UNOT} is a top view of the considered scenario, and all wire dipoles are orthogonal to the $xy$-plane ($z=0$). The obtained sum-rate is averaged over $10^3$ independent realizations for the locations of the scattering clusters and the loaded wire dipoles therein.

\ifsingle
\else
\begin{table}[t!]
\caption{Execution times $[s]$.}
\label{tab:time}
\centering
\resizebox{1\linewidth}{!}{%
\renewcommand{\arraystretch}{1}
\begin{tabular}{|c|c|c|c|c|c|}
\hline
\cellcolor[HTML]{EFEFEF} \textbf{$d$} & \textbf{Proposed alg.} & \textbf{BCD wMMSE} & \cellcolor[HTML]{EFEFEF} \textbf{$d$} & \textbf{Proposed alg.} & \textbf{BCD wMMSE} \\
\hline
 \cellcolor[HTML]{EFEFEF}  $\lambda/2$  & $1.17$ &  $4.84$ & \cellcolor[HTML]{EFEFEF} $\lambda/8$ & $19.48$  & $67.76$\\
\hline 
 \cellcolor[HTML]{EFEFEF} $\lambda/4$ & $2.52$ & $7.64$ & \cellcolor[HTML]{EFEFEF} $\lambda/16$ & $156.1$  & $1138$ \\
\hline 
\end{tabular}%
}
\renewcommand{\arraystretch}{1}
\end{table}
\fi
%
\change{We compare the performance of \name{} against two benchmark schemes, namely the method in~\cite{Abrardo2021}, which is denoted by BCD-wMMSE in the figures, and a \emph{mismatched} approach. The latter is obtained by assuming as objective function $\H_{\rme}$ in~\eqref{eq:NoInteractions}, which ignores the interactions between the RIS and the \glspl{eso}, and subsequenyly feeding the optimized $\W$ and $\Z_{\rmRIS}$ into the objective function in~\eqref{eq:Prob}.
In Fig.~\ref{fig:sumrate_N}, we show the average sum-rate versus the number of iterations for different values of $N$ and $d$, thus demonstrating the superior performance of \name{} both in terms of system sum rate and execution time to reach convergence, which is reported in Table.~\ref{tab:time}. In Fig.~\ref{fig:N_c}, we illustrate the average sum-rate versus the number of scattering clusters for different values of $N$ and $d$, demonstrating that ignoring the interactions between the RIS and the \glspl{eso} results in a significant degradation of the average sum-rate, especially as the number of \glspl{eso} increases. In Fig.~\ref{fig:K} we show the average sum-rate versus the number of \glspl{ue} $L$ for different values of $N$ and $d$. \name{} achieves higher performance for a small to moderate number of UEs, while the system becomes interference-limited for large $L$.}

\change{Finally, Fig.~\ref{fig:R0} shows the average sum-rate for small values of ${\rm{R}}_0$, assuming $N=64$ and $d=\lambda/8$. } 
In this case, the algorithm in~\cite{Abrardo2021} exhibits an unstable behavior. The reason is that the condition $\|\Deltab\G^i\|\ll 1$ for the Neumann series to be accurate is not included in the formulation of the optimization problem, whereas it is taken into account at each iteration of Algorithm~\ref{alg:A1}. This highlights the importance of adding the inequality constraint in~\eqref{eq:Prob2}, which makes the proposed approach robust-by-design when using the Neumann series approximation for small values of ${\rm{R}}_0$. For large values of ${\rm{R}}_0$, the two algorithms provide stable results, with the proposed algorithm converging faster.

\ifsingle
\else
\begin{figure}
    \begin{tabularx}{\linewidth}{XXX}
       \includegraphics[width=0.445\columnwidth]{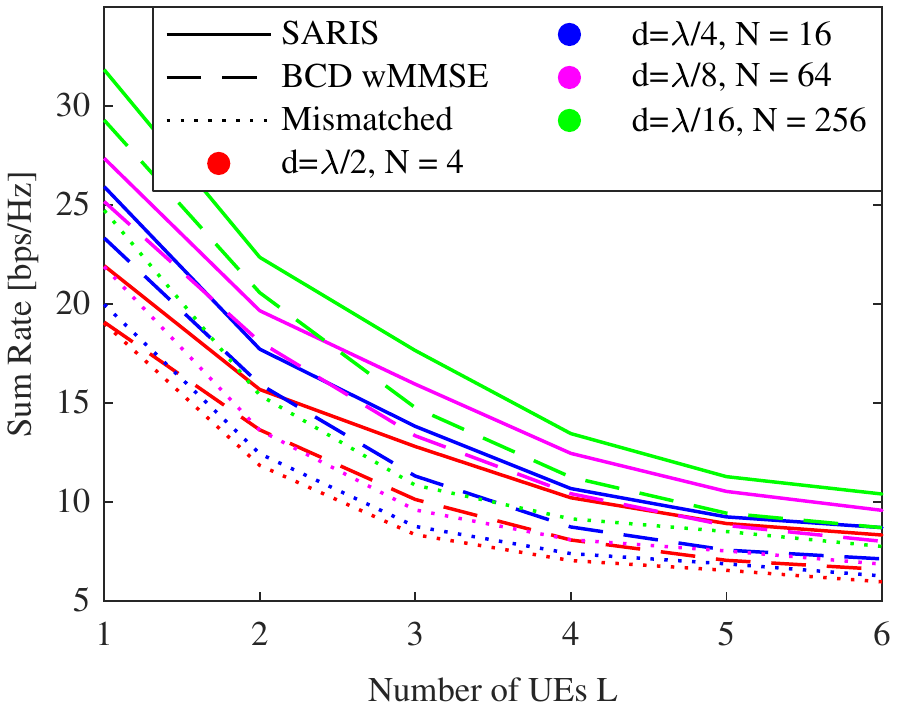}
       \vspace{-4mm}
   \caption{\change{Average sum-rate vs. $L$ and $d$.}} \label{fig:K}
    
    &
   
       \includegraphics[width=0.454\columnwidth]{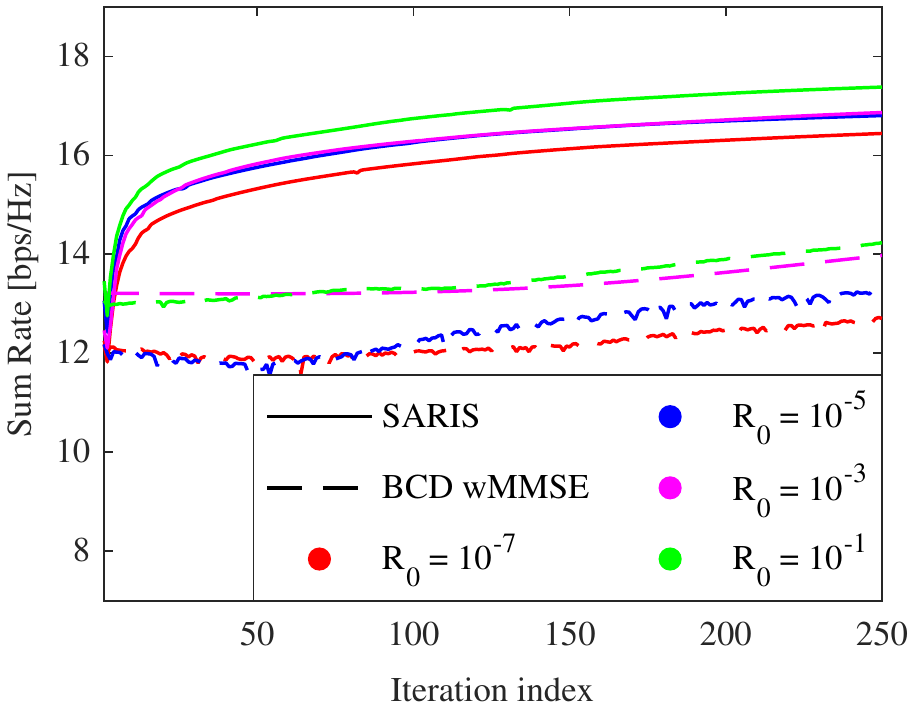}
       \vspace{-4mm}
       \caption{Convergence of Algo.~\ref{alg:A1} vs. $R_0$.} \label{fig:R0}
       \end{tabularx}
       \vspace{-5.5mm}
\end{figure}
\fi

\section{Conclusions}
In this letter, we have generalized a recently proposed model for RISs, based on mutually coupled loaded wire dipoles, to account for the presence of \glspl{eso}, i.e., the multipath from scattering objects. \change{The proposed approach relies on the DDA for scattering objects, and the interactions between the RIS elements and the \glspl{eso} are explictly taken into account. Based on the obtained end-to-end channel model, an efficient algorithm for optimizing the precoding at the transmitter and the configuration of the RIS, dubbed as \name{}, has been proposed. Numerical results demonstrate that it outperforms benchmark schemes available in the literature.}

\bibliographystyle{IEEEtran}
\bibliography{IEEEabrv,bibliography}

\end{document}